\documentclass[aps,prl,reprint,amsmath,amssymb]{revtex4-1}
\usepackage{graphicx}
\usepackage{hyperref}
\usepackage{pdfpages}

\newcommand*{\MyChi}{\raisebox{0.35ex}{\( \chi \)}}%

\def\pg{{\mathrm{pg}}} 
\def\sc{{\mathrm{sc}}} 
\def\U{U}
\def\sigvec{\pmb{\sigma}}

\begin{document}
\title{Signatures of pairing and spin-orbit coupling in correlation functions of Fermi gases}
\author{Chien-Te Wu, Brandon M. Anderson, Rufus Boyack and K. Levin}
\affiliation{James Franck Institute, University of Chicago, Chicago, Illinois 60637, USA}

\begin{abstract}
We derive expressions for spin and density correlation
functions in the (greatly enhanced)
pseudogap phase of spin-orbit coupled Fermi
superfluids.
Density-density correlation functions are found to be
relatively insensitive to the presence of these Rashba effects.
To arrive at
spin-spin correlation functions we derive new $f$-sum rules,
valid even in
the absence of a spin conservation law.
Our spin-spin correlation functions
are shown to be fully consistent with these $f$-sum rules.
Importantly, they provide
a clear signature of the Rashba band-structure and
separately help to establish the presence of a pseudogap.
\end{abstract}

\maketitle

\textit{Introduction.}$-$
Spin-orbit coupling (SOC) in superconductors and 
superfluids is a topic of much current interest~\cite{Kane2,dasSarma1,Sato}.
This is in large part because
there is some hope that (particularly in the presence of a
magnetic field) they may relate to the much sought after 
spinless $p_{x}+ip_{y}$ superfluid~\cite{ReadGreen}.
Two communities have united around
these issues: those working on cold Fermi superfluids with intrinsic
Rashba SOC~\cite{Zhaireview,GoldmanReview} and those studying 
superconductivity that is proximity induced
in a spin-orbit coupled material~\cite{Kane}.
To achieve this ultimate goal it is important to establish
that a given candidate for the 
$p_{x}+ip_{y}$ superfluid simultaneously exhibits
signatures of \textit{both} pairing and spin orbit coupling.
This would provide minimal evidence for a properly 
engineered ultracold atomic gas.
One therefore needs experimental signatures of these simultaneous
effects and this provides a central goal for the present paper. 

Here we address the signatures of this anomalous 
spin-orbit coupled superfluid 
as reflected in 
spin-spin and density-density
correlation functions.
Our work builds on the observation 
~\cite{Fermions1,Fermions2,Fermions3,Fermions4,Fermions5,Fermions6}
that in the presence of Rashba
SOC, pairing (in the form of pseudogap
effects~\cite{ourreview,JS2}) is significantly enhanced.
For this reason (and because the correlation functions
are free of the complications of collective mode effects)
we focus here on the normal phase. 
We show how even without condensation,
the frequency dependent spin response 
exhibits features which relate to the Rashba ring band-structure,
as well as to the presence of a pairing gap. In 
contrast, the density response is relatively unaffected by SOC. 
Previous work has focused on identifying SOC without~\cite{Zwierlein,Pengjun} or with pairing~\cite{SOCrf}
via the  
one body spectral function. As compared with
Ref.~\cite{SOCrf}, we find less subtle
features in the two particle response.
At the very least the spin correlation functions provide
complementary and accessible (via neutrons or
two photon Bragg scattering~\cite{ValePRL08}) information.

Validating any theory of correlation functions requires
satisfying important constraints~\cite{Rufuspaper}.
Indeed, the absence of conservation laws  
complicates
all spin transport in spin-orbit coupled materials. 
Thus, it is extremely important to find underlying principles for
establishing self consistency. 
To address this issue, here we use the Heisenberg equations
of motion to derive $f$-sum rules 
for the spin-spin correlation functions, which also
provides important constraints on
our numerical calculations.
While a magnetic field is necessary for arriving at topological
order, we begin by ignoring this additional complication.
There is a substantial literature investigating correlation functions
in the superfluid phase (without pseudogap
effects) which we note here~\cite{Kitagawa,RoyHall,Kallin,Yakovenko}.

In this Rapid Communication we present two main results. 
The first is a consistent derivation of spin-spin correlation functions in spin-orbit coupled Fermi gases. The second establishes
qualitative experimental signatures reflecting separately the
presence of a pairing gap and
of SOC.
Readers interested in the experimental signatures need only a cursory exploration of the mathematical derivation that precedes it.

\begin{figure}[t]
\includegraphics[width=8.6cm]{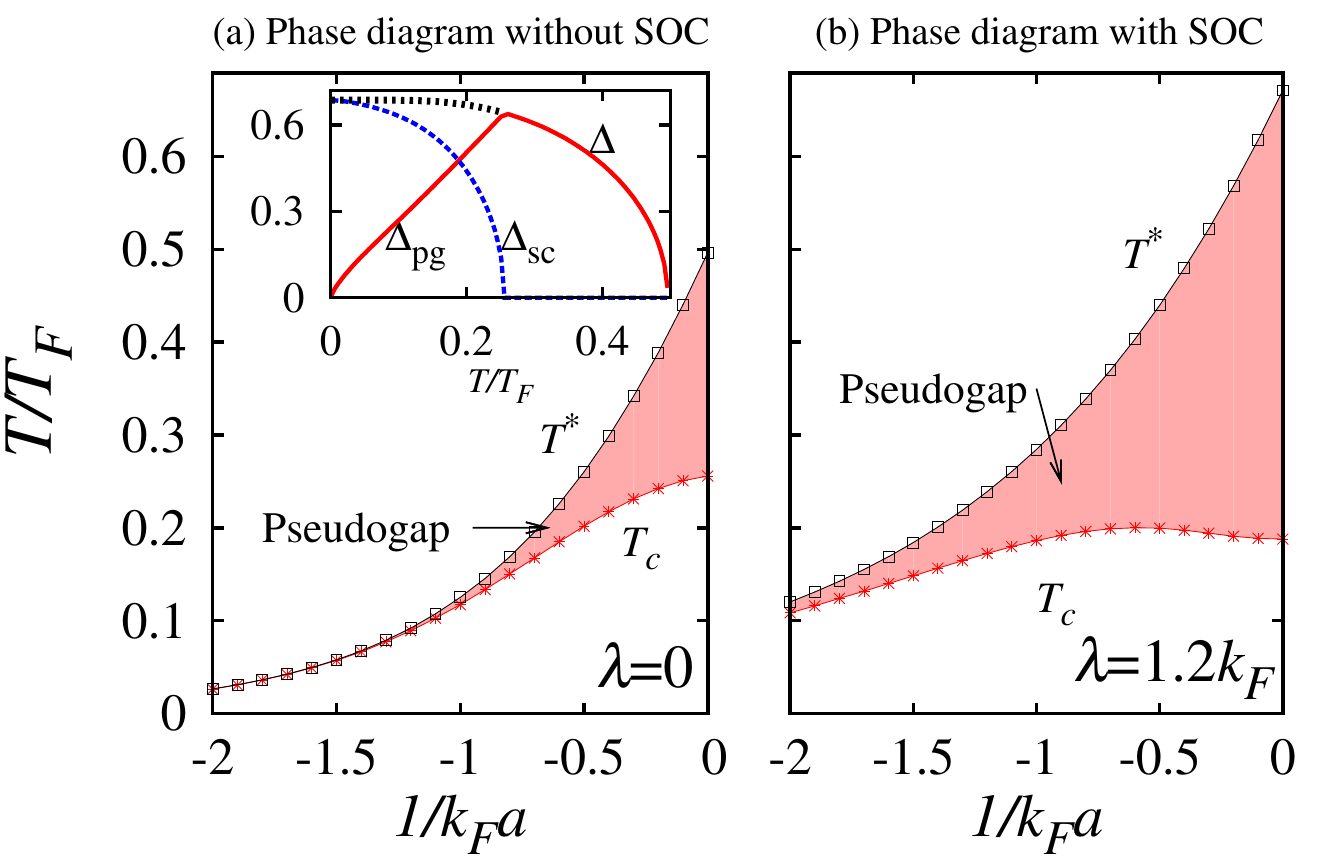}
\caption{Phase diagrams for a degenerate fermi gas without (a) and with 
(b) SOC. Plotted is $T^*$ as a function of inverse scattering length, 
$1/k_Fa$, indicating where pairing 
first sets in, while  
$T_c$ marks the onset of condensation to
the superfluid phase. The plot shows the weakly interacting regime
with $1/k_F a<0$.  As spin-orbit coupling is turned on, pairing is enhanced 
leading to a larger pseudogap region. The inset in (a) indicates the
temperature dependences of the component gap parameters (defined
in the text) at unitarity, $1/k_F a = 0$. }
\label{fig:phaseplot}
\end{figure}

\textit{Background Theory.}$-$
We consider a gas of fermions whose single particle Hamiltonian is 
$H^0({\bf k}) = k^2/2m - \mu + \lambda \sigvec \cdot {\bf k}_\perp/m$
for a particle of mass $m$, spin-orbit coupling momentum $\lambda$, momentum ${\bf k} = (k_x, k_y, k_z)$, 
in-plane momentum ${\bf k}_\perp = (k_x, k_y, 0)$, and vector of Pauli matrices $\sigvec = (\sigma_x, \sigma_y, \sigma_z)$. 
Throughout this paper we set $\hbar=k_B=1$.
To describe our spin-orbit coupled Fermi gas with pairing, we use a $4\times4$ inverse Nambu Green's function 
\begin{equation}
\label{eq:GF}
\mathcal{G}^{-1}(K) 
=
\begin{pmatrix}
G_{0}^{-1}(K) & \Delta\\
\Delta & \widetilde{G}_{0}^{-1}(K)
\end{pmatrix},
\end{equation}
that acts on the spinor $\Psi_{\bf k}^T = \left( c_{{\bf k} \uparrow}, c_{{\bf k} \downarrow }, -c^\dagger_{-{\bf k} \downarrow}, c^\dagger_{-{\bf k}\uparrow} \right)$ for a fermion annihilation (creation) operator $c_{{\bf k}s} (c_{{\bf k}s}^\dagger)$ of spin $s=\uparrow,\downarrow$ and momentum ${\bf k}$. 
In the Green's function $\Delta$ is a pairing gap, the 4-vector $K=(i\omega,\mathbf{k})$ with Matsubara frequency $i \omega$, and the non-interacting inverse particle Green's function is 
$G_{0}^{-1}(K) = i\omega - H^0({\bf k})$ and hole Green's function $\widetilde{G}_{0}^{-1}(K)=i\sigma_y [G_{0}^{-1}(-K)]^T i\sigma_y = i\omega + H^0({\bf k})$.

This superfluid can be studied at the mean field 
level~\cite{Fermions1,Fermions2,Fermions3,Fermions4,Fermions5,Fermions6},
where one has the usual gap equation:
\begin{equation}\label{eq:GE}
1=\frac{g}{2}\sum_{K}\mathrm{Tr}\left[G(K)\widetilde{G}_{0}(K)\right],
\end{equation}
where $\sum_K = T\sum_{i\omega} \sum_{\bf k}$ is a sum over momentum and Matsubara frequencies at temperature $T$, and where the many-body Green's function $G(K)$ is found from the inverse of Eq.~(\ref{eq:GF})
\begin{eqnarray}\label{eq:BGF} 
G^{-1}(K) -G_{0}^{-1}(K) & = & -\Sigma(K) = - \Delta^2\widetilde{G}_{0}(K).
\end{eqnarray}
Except for the matrix structure in the above equations, these are the
usual definitions of the fermionic Green's functions and self energy
associated with BCS theory; the gap equation also appears
in the literature~\cite{Fermions1,Fermions2,Fermions3,Fermions4,Fermions5,Fermions6}. 
As in the cold atoms literature, we will regularize the gap equation by replacing the interaction strength $g$ with the scattering length $a$ through
 $1/g = m/4\pi a - 1/V \sum_\textbf{k} m/\textbf{k}^2 $.

We now want to include pair fluctuations, or pseudogap
(pg) effects, in a fashion fully consistent with 
both the ground state and the
mean field equations that have been extensively
studied in previous work~\cite{Fermions1,Fermions2,Fermions3,Fermions4,Fermions5,Fermions6}.
The approach we outline below was introduced in the
context of high temperature superconductors~\cite{ourreview,Chen2},
but it has also been applied to spin-orbit coupled superfluids~\cite{GGO}.

Pair fluctuations lower the phase transition temperature $T_c$ 
relative to its mean field value denoted by $T^*$. 
The latter is the temperature at which the pairing gap, 
determined by Eq.~(\ref{eq:GE}), first becomes non zero.  
(Note that $T^*$ does not reflect a broken symmetry state
and is not a true phase transition).
As a result, a central component of the present theory is that
the usual Thouless condition (on the $t$-matrix) for the 
instability of the normal phase~\cite{S64} must be
modified to include a  well developed excitation gap.

Imposing consistency with mean field theory
for the pairing gap obtained from Eq.~(\ref{eq:GE}) leads to a modified
Thouless 
condition: 
\begin{equation}\label{eq:TPG}
t_{\pg}(Q)\equiv\frac{g}{1+g\MyChi(Q)}\rightarrow\infty,\text{\ as\ } Q\rightarrow 0,
\end{equation}
where
\begin{equation}
\MyChi(Q) \equiv - \frac{1}{2}\sum_{K}\mathrm{Tr}\left[G(K)
\widetilde{G}_{0}(K-Q)\right].
\end{equation}
In this way, above $T_c$ the square of the
excitation gap of the non-condensed
pairs, called $\Delta_{\pg}^2$, is to be
associated with $\Delta^2$,
given by Eq.~(\ref{eq:GE}) 
\footnote{It is possible to consider more precise $t$-matrix based numerical [see J. Maly et al, Physica C 321, 113 (1999)] and analytic [He et al, PRB 76, 224516 (2007)] schemes; as long as the pseudogap vanishes at the mean field $T^*$, there is no relevant qualitative difference.}.
Establishing $T_c$, however, requires that we find
a constraint on $\Delta_{\pg}^2$ applicable at and below $T_c$.
Indeed, once $T < T_c$, there must be another
contribution to the self energy $\Sigma(K)$ involving 
$Q=0$ or superconducting (sc) pairs. Then  
the self energy of Eq.~(\ref{eq:BGF}) can be written 
in terms of a $t$-matrix (as in the Thouless condition) 
$\Sigma(K)= - \sum_Q t(Q) \widetilde{G}_0 (K-Q),$
\cite{Chen2}: $ t_{\pg}(Q)+t_{\sc}(Q)$, where
$t_{\sc}(Q)\equiv-\frac{\Delta_{\sc}^2}{T} \delta(Q)$ represents
the condensate. As a consequence 
$\Sigma(K)=\Sigma_{\sc}(K) + \Sigma_{\pg}(K)$.   
Because of Eq.~(\ref{eq:TPG}), the  
quantity $t_{\pg}(Q)$ is strongly peaked at $Q=0$ and
$\Sigma_{\pg}(K)\approx\Delta_{\pg}^2 \widetilde{G}_0(K)$.

This relation implies that $\Delta_{\pg}^2 = - \sum_{Q \neq 0} t_{\pg}(Q)$,
or equivalently, the pseudogap is approximated as a thermal gas of composite bosons. 
The transition temperature $T_c$ is determined as the temperature at which 
this value of $\Delta_{\pg}$, when the transition
is approached from below, 
intersects with the mean field gap equation, which corresponds to
$\Delta_{\pg}$ above $T_c$ in the normal state.
In this way, in the ordered phase the total self energy
$\Sigma(K)=\Delta^{2}\widetilde{G}_{0}(K)$,
where $\Delta^2 = \Delta_{\sc}^2 + \Delta_{\pg}^2$.
In the inset of Fig.~\ref{fig:phaseplot}(a), we plot the temperature dependence
for both $\Delta_{\sc}$ and $\Delta_{\pg}$, as well as the total
excitation gap $\Delta$.

We emphasize that a central distinguishing feature of the present
approach is that $T_c$ is determined in the presence of a well
developed gap at $T_c$. This contrasts with the scheme of Nozieres and
Schmitt-Rink~\cite{NSR} and is similarly different from path
integral-collective mode schemes~\cite{superden1}.
The latter introduce Goldstone
bosons, but importantly these do not
renormalize the mean field transition temperature 
which remains at $T^*$.

\begin{figure*}
\includegraphics[width=2.25in,clip]
{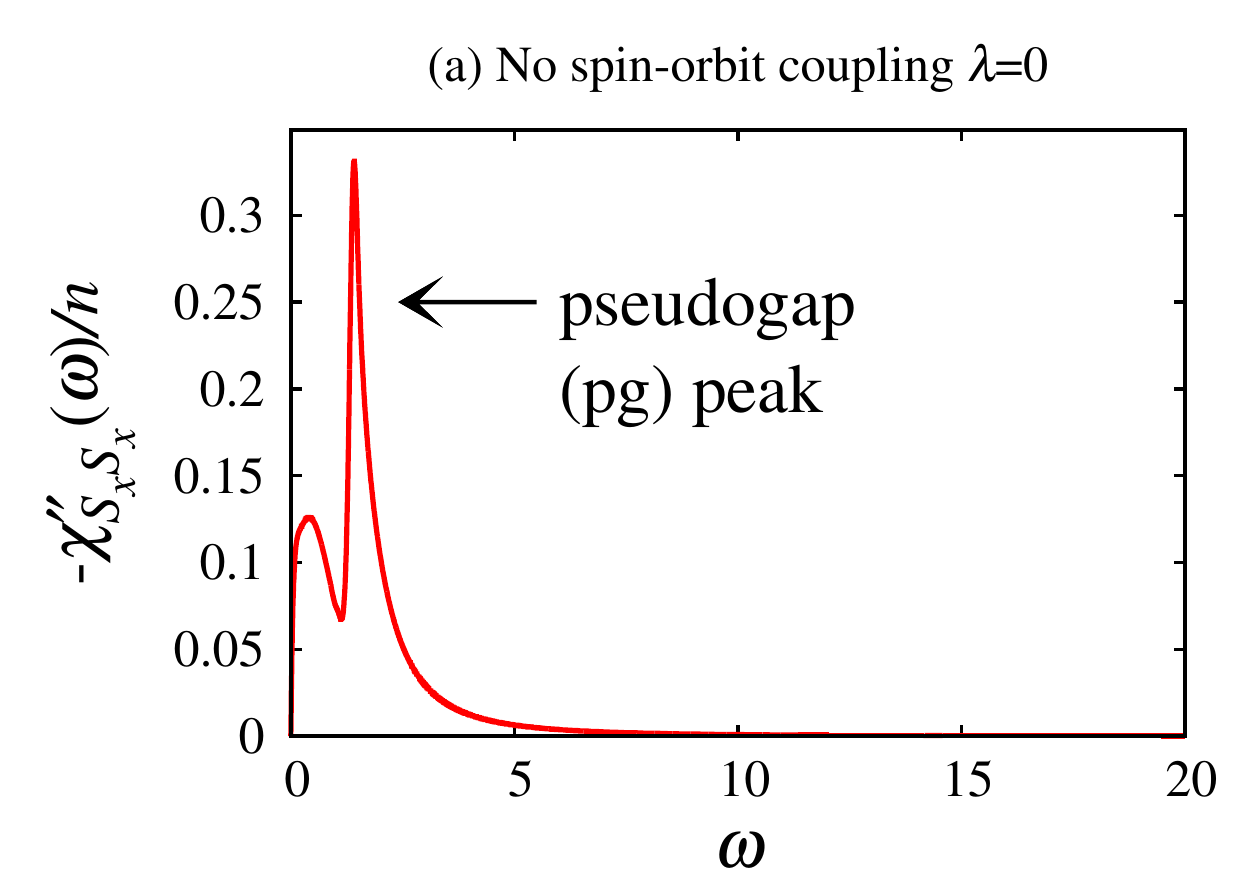}
\includegraphics[width=2.25in,clip]
{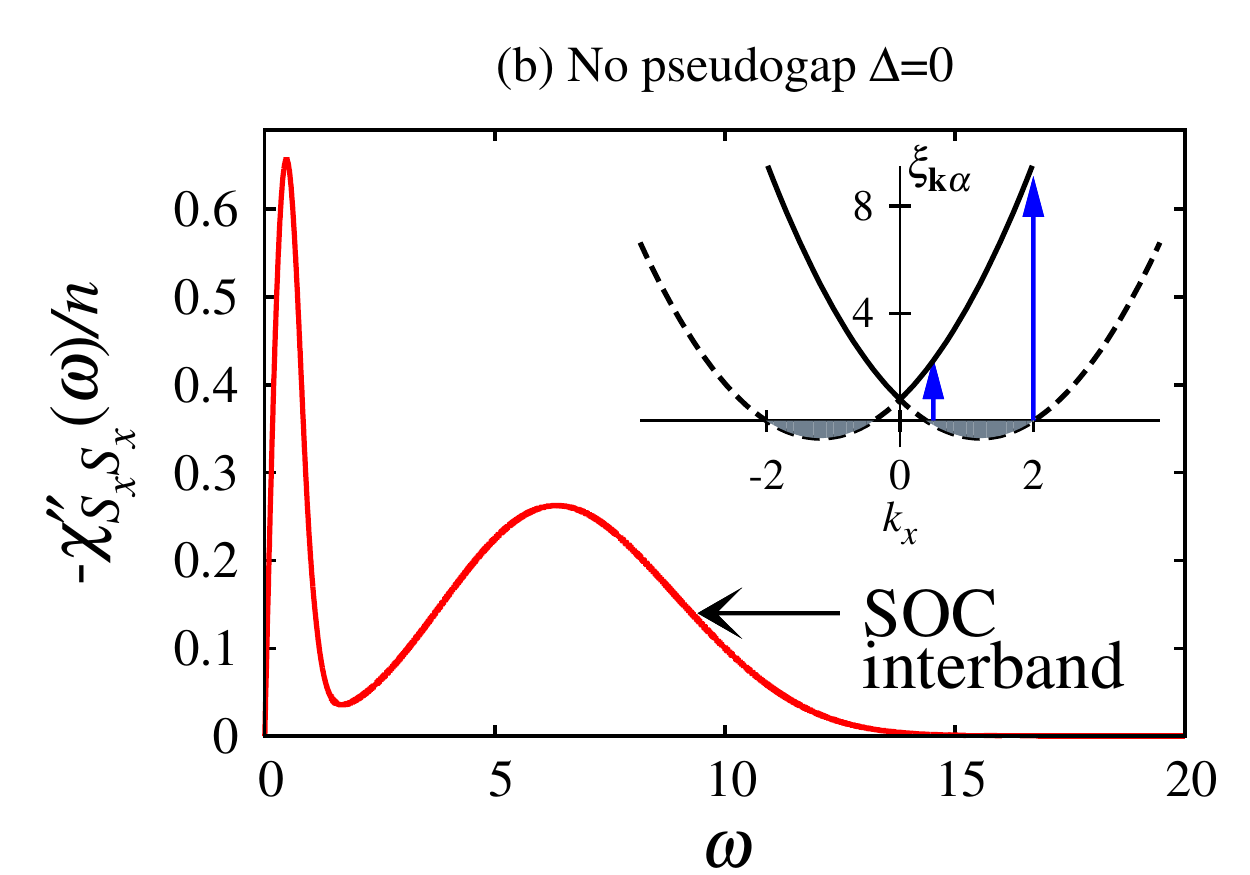}
\includegraphics[width=2.25in,clip]
{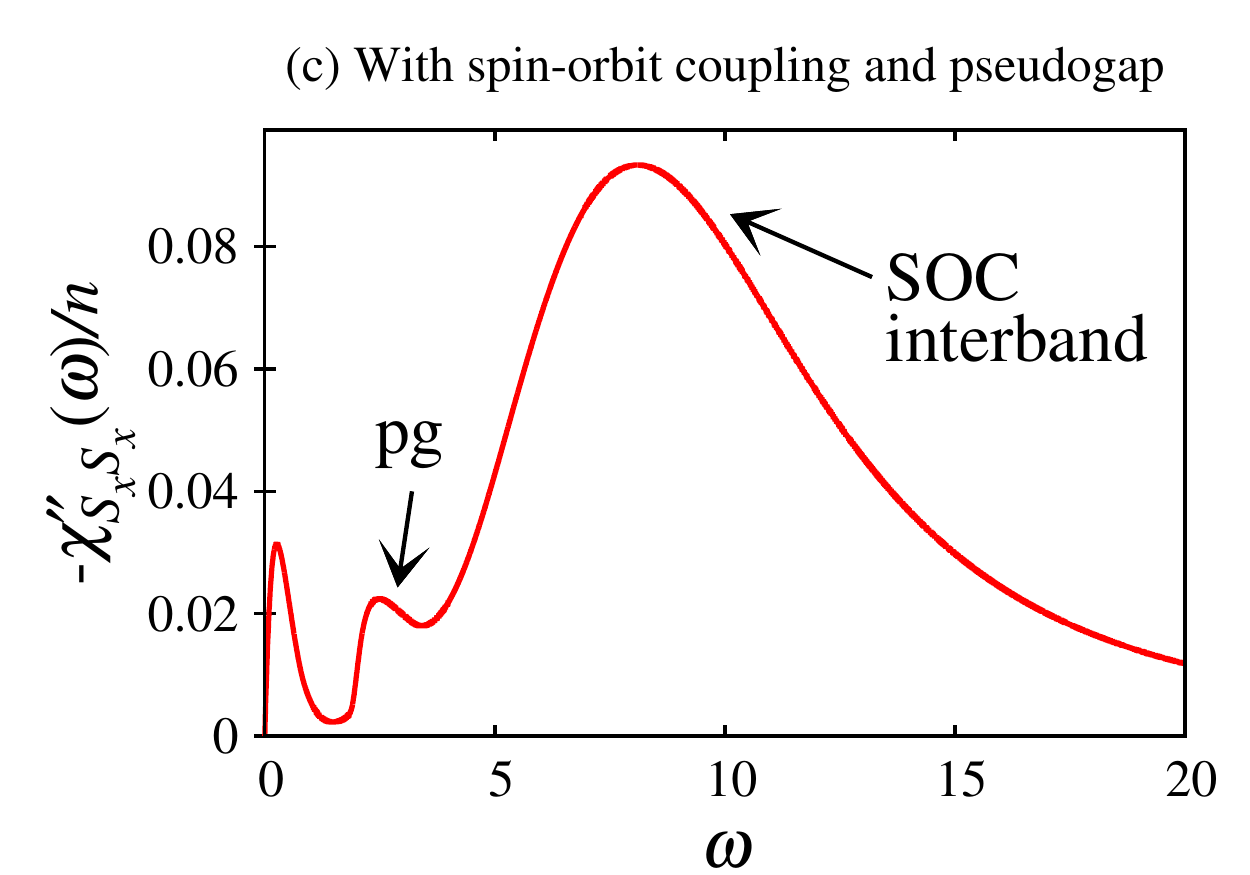}
\caption{The spin-spin response function as a function of
frequency.
Figure~\ref{fig:chiss}(a) corresponds to a pseudogap phase without SOC. The
peak at lower $\omega$ reflects thermally excited fermions while
the second peak is at $\omega$ comparable to the pseudogap
energy scale, where pairs are now broken.
Since $\lambda = 0$,
the spin-spin and density-density correlation functions
are the same.
In Fig.~\ref{fig:chiss}(b) the pseudogap is set to zero with fixed $\lambda =1.2 k_F$.
The lower energy intra-helicity band peak is enhanced and
one sees an SOC-related peak. 
As shown in the inset, the
low-energy threshold reflects the onset of inter-helicity
band transitions, while the high-energy endpoint occurs when
these transitions are no longer possible.
In Fig.~\ref{fig:chiss}(c), in the presence of both a pseudogap and SOC, 
one sees a combination of the effects in the previous two panels.
Spectral weight is transferred to yield
a larger SOC peak, reflecting the gapping of the low
energy contributions.
All quantities are measured relative to the Fermi energy, $E_F$, or 
Fermi momentum $k_F$.
}
\label{fig:chiss}
\end{figure*}

Figure~\ref{fig:phaseplot} shows the calculated phase diagram, 
plotting $T_c$ and $T^*$, in the 
absence (a) and presence (b) 
of Rashba SOC. The latter is in reasonable agreement with
the results of Ref.~\cite{GGO}. 
We restrict these plots to the weak pairing side of resonance.
Here  we observe a greatly enhanced pseudogap regime denoted by an enhancement of $T^*$ without significant enhancement of $T_c$.
The behavior of $T^*$ has been attributed
to the enhancement of the pairing 
attraction~\cite{Fermions1,Fermions2,Fermions3,Fermions4,Fermions5,Fermions6}, due to an
increased density of states near the minimum of the Rashba ring.
Since $T_c$ is obtained in the presence
of a gap at $T_c$, stronger pairing (reflected in $T^*$)
is offset by an increasingly gapped density of states.
This leads to a relatively constant $T_c$ as a function
of interaction strength.

\textit{Density/current and $f$-sum rules.}$-$
To characterize the anomalous, normal, and superfluid phases in more detail,
we investigate both the density/current and spin correlation functions, considering the former first.
For systems with a $\U(1)$ symmetry, the Ward-Takahashi
identity (WTI) provides an important constraint on the full vertex $\Gamma^{\mu}(\widetilde{K},K)$ which enters
into the correlation functions. Given a mean field like self energy,
it is possible to analytically solve the WTI, and obtain the full vertex function
along with the full correlation function~\cite{S64,Rufuspaper}.

We define the generalized correlation function
\begin{equation}\label{eq:CF}
P^{\mu\nu}(Q)=\sum_{K}\mathrm{Tr}\left[G(\widetilde{K})\Gamma^{\mu}(\widetilde{K},K)G(K)\gamma^{\nu}(K,\widetilde{K})\right],
\end{equation}
where $\widetilde{K}\equiv K+Q$ and $\gamma^{\mu}(\widetilde{K},K)$ is a bare vertex. From this we have the density-density $\MyChi_{\rho\rho}(Q)\equiv P^{00}(Q)$ and current-current correlation functions $\tensor{\MyChi}_{JJ}(Q)\equiv P^{ij}(Q),i,j\in\{1,2,3\}$.
The bare and full vertices satisfy respectively
\begin{eqnarray} \label{eq:GammaMu}
q_{\mu}\gamma^{\mu}(\widetilde{K},K) & = & G_0^{-1}(\widetilde{K})-G_0^{-1}(K), \label{eq:baregamma}\\
q_{\mu}\Gamma^{\mu}(\widetilde{K},K) & = & G^{-1}(\widetilde{K})-G^{-1}(K), \label{eq:fullgamma}
\end{eqnarray}
with the latter a consequence of the WTI. 
We now specialize to systems with the self energy as in 
Eq.~(\ref{eq:BGF}). Using the WTI above $T_{c}$ we have
\begin{equation}\label{eq:FV}
\Gamma^{\mu}(\widetilde{K},K)=\gamma^{\mu}(\widetilde{K},K)+
\Delta^{2}\widetilde{G}_{0}(\widetilde{K})\widetilde{\gamma}^{\mu}(\widetilde{K},K)\widetilde{G}_{0}(K),
\end{equation}
where $\widetilde{\gamma}^{\mu}(\widetilde{K},K) = \sigma_y {\gamma}^{\mu}(-\widetilde{K},-K)^T \sigma_y$ is a time-reversed vertex.
Inserting the full vertex into Eq.~(\ref{eq:CF}) then gives the correlation functions above $T_{c}$.

One can incorporate superconducting (or equivalently superfluid) terms within
this formalism building on Eq.~(\ref{eq:FV}) and, for example, 
address the superfluid density~\cite{Chen2}, as outlined in the supplement. One 
considers the transverse response 
$P_{T}^{\mu\nu}(Q)$ which contains no collective modes:
\begin{eqnarray}\label{eq:PRR}
P_{T}^{\mu\nu}(Q)&=&\sum_{K}
\mathrm{Tr}\biggl\{\left[G(\widetilde{K})\gamma^{\mu}(\widetilde{K},K)G(K)\right.
\nonumber\\&+&{F}_{\pg}(\widetilde{K})\widetilde{\gamma}^{\mu}(\widetilde{K},K)
\widetilde{{F}}_{\pg}(K)
\nonumber\\&-&\left.F_{\sc}(\widetilde{K})\widetilde{\gamma}^{\mu}(\widetilde{K},K)
\widetilde{F}_{\sc}(K)\right]
\gamma^{\nu}(K,\widetilde{K})\biggr\},
\end{eqnarray}
where $F_{m}(K) = \Delta_{m} \widetilde{G}_{0}(K)G(K) = \widetilde{F}_{m}(K)$ for $m\in\{\sc,\pg\}$. 
Note that $F_{\pg}$ does not represent an anomalous Green's function, but rather reflects a
vertex correction to the correlation functions~\cite{Rufuspaper}.
There is a disagreement in the literature~\cite{Kitagawa,Fermions4}
as to the importance of collective modes in the spin response below $T_c$. We agree with the results
of Ref.~\cite{Kitagawa}, where collective modes were included.

As shown in the supplementary material,
when one integrates over the entire frequency range, a consequence of
the WTI is that the $f$-sum rule is satisfied: 
\begin{equation}\label{eq:FSR}
\int \frac{d\omega}{\pi}\left(-\omega\MyChi^{\prime\prime}_{\rho\rho}(\omega,{\bf q})\right)
=\frac{nq^2}{m},
\end{equation}
where $\MyChi^{\prime\prime}_{\rho\rho}$ is the imaginary part of the density response function.
This $f$-sum rule depends on the total particle number $n$ and the bare mass $m$. Since $\lambda$ does not enter, the presence of spin-orbit coupling does not modify the weight of the $f$-sum rule.

\textit{Spin response and $f$-sum rules.}$-$
In the spin channel, where there is no $\U(1)$ symmetry to justify
the use of the WTI. Nevertheless, we are able to provide
an
{\em a posteriori} check on any proposed correlation function via
a sum rule which we now derive. We define
$\MyChi_{S_{i}S_{j}}\left(i\omega,\mathbf{q}\right)\equiv \int d\tau \,e^{i\omega \tau} \left\langle T_\tau S_{\mathbf{q}i}\left(\tau\right) S_{-\mathbf{q}j}\left(0\right)\right\rangle$
where $T_\tau$ is the time ordering operator and 
$S_{\mathbf{q} i}=\sum_{\mathbf{k}ss^\prime}c_{\mathbf{k}s}^{\dagger}\left(\sigma_{i}\right)_{ss^\prime}c_{\mathbf{k}+\mathbf{q}s^\prime}$
is the many-body spin density operator.
Using the Heisenberg equations of motion and the properties of Fourier transforms, the sum rule for the spin-spin correlation function $\MyChi^{\prime\prime}_{S_{i}S_{j}}$ can be shown to be
\begin{eqnarray}\label{eq:SSR1}
\int \frac{d\omega}{\pi} \left(-\omega\MyChi^{\prime\prime}_{S_{i}S_{j}}\left(\omega,\mathbf{q}\right)\right)
& = & \left\langle \left[\left[\mathcal{H}_0,S_{\mathbf{q}i}\right],S_{-\mathbf{q}j} \right]\right\rangle,
\end{eqnarray}
where $\mathcal{H}_{0}=\sum_{ss^\prime\mathbf{k}}c_{\mathbf{k}s}^{\dagger}H^{0}_{ss^\prime}\left(\mathbf{k}\right)c_{\mathbf{k}s^\prime}$ and $\MyChi^{\prime\prime}_{S_{i}S_{j}}$ is the singular part of $\MyChi_{S_{i}S_{j}}$ found by analytically continuing $i\omega \rightarrow \omega + i\delta$ and then taking the $\delta \rightarrow 0$ limit.

Here we give the explicit result, for two example cases of interest
and present further details in the supplementary material: 
\begin{align}\label{eq:SSR}
\int \frac{d\omega}{\pi}\left(-\omega\MyChi^{\prime\prime}_{S_{i}S_{i}}\left(\omega,\mathbf{q}\right)\right)
&=\frac{nq^{2}}{m}-\frac{4\lambda}{m}\sum_{\mathbf{k}\alpha}\alpha f_{ii}
n_{\mathbf{k}\alpha}.
\end{align}
where $i\in\{x,z\}$, $f_{zz}=k_{\perp}$, $f_{xx}=k_{x}^{2}/k_{\perp}$ 
and $n_{\mathbf{k}\alpha}=T\sum_{i\omega}G^{\alpha}_{H}(K),$ with $G^{\alpha}_{H}(K)$ a
helicity Green's functions to be defined in the next section.

\textit{Correlation functions in the helicity basis.}$-$
In the absence of a magnetic field, helicity is a good quantum number, and the correlation functions are
most easily expressed in terms of the helicity Green's functions~\cite{GGO}:
\begin{eqnarray}
G_{H}^{\alpha}(K)&\equiv&\frac{u_{\mathbf{k}\alpha}^2}{i\omega
-E_{\mathbf{k}\alpha}}+\frac{v_{\mathbf{k}\alpha}^2}{i\omega+E_{\mathbf{k}\alpha}},\\
{F}_{H}^{\alpha}(K)&\equiv& u_{\mathbf{k}\alpha}v_{\mathbf{k}\alpha}
\left(\frac{1}{i\omega+E_{\mathbf{k}\alpha}} 
- \frac{1}{i\omega-E_{\mathbf{k}\alpha}}\right),
\end{eqnarray}
where $E_{\mathbf{k}\alpha}=\sqrt{\xi^2_{\mathbf{k}\alpha} + \Delta^2}$, 
$\xi_{\mathbf{k}\alpha} = k^2/2m - \mu + \alpha\lambda k_{\perp} /m$ is an eigenvalue of $H^0({\bf k})$, 
and ${F}^{\alpha}_{H}(K)$ represents the pseudogap, or equivalently vertex contribution. Here 
$\alpha=\pm$ denotes the helicity index and the coherence factors satisfy
$u_{\mathbf{k}\alpha}^2=\tfrac{1}{2}(1+\xi_{\mathbf{k}\alpha}/E_{\mathbf{k}\alpha})$,
$u_{\mathbf{k}\alpha}^2+v_{\mathbf{k}\alpha}^2=1.$

It follows from the vertex function in Eq.~(\ref{eq:FV}) that
the explicit form for the $f$-sum rule~\cite{Rufuspaper} consistent 
density-density correlation function is
\begin{align}&\MyChi_{\rho\rho}(\omega,{\bf q})=\frac{1}{2}\sum_{K,\alpha,
\alpha^{\prime}}\left(1+\alpha\alpha^{\prime}\cos\left(\phi_{\bf{k+q}}-\phi_{{\bf k
}}\right)\right)\nonumber\\&\times\left[G_{H}^{\alpha}(K)G_{H}^{\alpha^{\prime}
}(\widetilde{K})+{F}_{H}^{\alpha}(K){F}_{H}^{\alpha^{\prime}}(\widetilde{K})\right].
\label{eq:17}\end{align}
The angle $\exp(i\phi_{\mathbf{k}})=(k_{x}+ik_{y})/k_{\perp}$,
so that $\exp(i\phi_{-\mathbf{k}})=-\exp(i\phi_{\mathbf{k}}).$

The spin-spin correlation functions are constructed using
their form 
below $T_{c}$ (deduced using the path integral~\cite{Kitagawa})
with appropriate sign changes in the pseudogap relative to the
condensate gap. These sign changes, which appear in the $T< T_c$
Ward-Takahashi identity~\cite{S64}
are essential for satisfying sum
rules.

As can be shown, in the normal phase
the following expression for the spin-spin correlation functions are fully compatible with the spin $f$-sum rules given in Eq.~(\ref{eq:SSR}):
\begin{align}
&\MyChi_{S_{i}S_{i}}(\omega,{\bf q})=\frac{1}{2}\sum_{K,\alpha,\alpha^{\prime}}\left(1\pm\alpha\alpha^{\prime}\cos\left(\phi_{{\bf k+q}}\pm\phi_{{\bf k}}\right)\right)\nonumber\\&\times\left[G_{H}^{\alpha}(K)G_{H}^{\alpha^{\prime}}(\widetilde{K})+{F}_{H}^{\alpha}(K){F}_{H}^{\alpha^{\prime}}(\widetilde{K})\right],
\end{align}
where the $+,-$signs are for $\MyChi_{S_{x}S_{x}},\MyChi_{S_{z}S_{z}}$ respectively.

\textit{Numerical Results.}$-$
We now look for qualitative new physics in the spin-spin response functions. 
We numerically calculate the response function
$\MyChi^{\prime\prime}_{S_xS_x}({\bf q},\omega)$
at fixed ${\bf q}=(0.5,0,0) k_F$ as a function of $\omega$
~\footnote{We introduce a finite life time $\Gamma=0.05$ in the self-energy both to distinguish this term from the condensate and for numerical stability.},
and for definiteness consider $T=0.28 T_F>T_c$ and unitary
scattering, $1/k_F a=0$. We plot the results in Fig.~\ref{fig:chiss}. In order to illustrate the
physics, in Fig.~\ref{fig:chiss}(a) and Fig.~\ref{fig:chiss}(b),
Rashba SOC or pseudogap effects
were set to zero respectively, while Fig.~\ref{fig:chiss}(c) shows their
combined effects. 
The $f$-sum rules derived above are important for
constraining numerical results of the spin-spin and
density-density correlation functions.
Comparison between our numerical calculations and the exact $f$-sum rules
agreed to within a few percent.

In Fig.~\ref{fig:chiss}(a) we set $\lambda=0$.
In this case, above $T_c$ the spin and density correlations are equal
($\MyChi^{\prime\prime}_{S_xS_x}=\MyChi^{\prime\prime}_{\rho\rho}$) and this
function is plotted in the figure. Two low energy peaks are observable, as found
in our earlier work~\cite{Hao_prl}.
The lower frequency peak reflects contributions from
thermally excited fermions, while the higher frequency peak is associated with
the contribution from broken pairs which appears at a
threshold associated with the pseudogap.

In Fig.~\ref{fig:chiss}(b) we set $\Delta_{\pg}=0$ and plot 
$\MyChi^{\prime\prime}_{S_xS_x}$ for a pure SOC system with $\lambda=1.2 k_F$.
(We do not show $\MyChi^{\prime\prime}_{\rho\rho}$ since there is still no qualitative 
signature of $\lambda \neq 0$.)
The response $\MyChi^{\prime\prime}_{S_xS_x}$ shows two peaks, but one is at a considerably 
higher energy compared to Fig.~\ref{fig:chiss}(a). The lower frequency peak reflects
intra-helicity band contributions while the
larger frequency peak is due to inter-helicity effects.

Importantly, this figure shows how
the physics of the Rashba ring band-structure can be directly probed by
the spin-spin response function. To illustrate this, in the inset we plot
the dispersion relation of two helicity bands.
The horizontal line denotes the self-consistently determined chemical potential, 
chosen so that occupied fermions mostly reside
in the Rashba ring. The onset of the inter-helicity band transition 
energy is given by the energy difference between two bands 
positioned on the inner circle of the ring, while the endpoint
frequency for this peak
is determined by the 
outer circle. These energy differences roughly match the width
observed in the high frequency peak in the main plot. (The smearing
of the width is because we have a non-zero momentum
${\bf q}$ and $T\neq0$.)

Finally, in Fig.~\ref{fig:chiss}(c) we 
plot $\MyChi^{\prime\prime}_{S_xS_x}$
for the case where both pseudogap
and Rashba SOC are present.
Here we observe three distinct peaks. The first is associated with
thermally excited fermions within the lowest helicity band,
the second with the breaking of the preformed (pg) pairs and the third
mainly with the inter-helicity transitions discussed in the previous
panel. We also observe some inter-play between pairing 
and the high frequency SOC peak, as 
this inter-helicity band peak
is pushed toward slightly higher energies.

\textit{Conclusion.}$-$
A major finding of this paper is that spin-spin correlations provide a clear signature of the simultaneous presence of Rashba modified band-structure and of a pairing gap. Signatures of both are a necessary (but clearly not sufficient) condition for ultimately obtaining a topological superfluid. This should complement observations which are based on the single particle response functions in different experiments either in cold gases~\cite{SOCrf,Zwierlein,Pengjun} or in condensed matter. Our spin correlation functions are consistent with sum rules which we derive in this paper. These provide important constraints on the spin response which is complicated by the fact that spin conservation laws are unavailable for spin-orbit coupled systems.

\textit{Acknowledgements.}$-$
This work was supported by NSF- DMR-MRSEC 1420709. We are grateful to P. Scherpelz and A. Sommer for helpful conversations.
\bibliography{Review}
\clearpage


\section*{Supplementary material}

# Supplementary Material: Signatures of pairing and spin-orbit coupling in correlation functions of Fermi gases

Chien-Te Wu, Brandon M. Anderson, Rufus Boyack and K. Levin
*James Franck Institute, University of Chicago, Chicago, Illinois 60637, USA*

## I. GREEN'S FUNCTIONS AND SELF ENERGY IN THE HELICITY BASIS

In evaluating the correlation functions, it is convenient to have the Green's functions written in the helicity basis. Here we present a brief discussion of the simple expressions used in the main text:

$$G_H^\alpha(K) \equiv \frac{u_{\mathbf{k}\alpha}^2}{i\omega - E_{\mathbf{k}\alpha}} + \frac{v_{\mathbf{k}\alpha}^2}{i\omega + E_{\mathbf{k}\alpha}}, \tag{1}$$

$$F_H^\alpha(K) \equiv u_{\mathbf{k}\alpha} v_{\mathbf{k}\alpha} \left( \frac{1}{i\omega + E_{\mathbf{k}\alpha}} - \frac{1}{i\omega - E_{\mathbf{k}\alpha}} \right), \tag{2}$$

where the Bogoliubov spectrum $E_{\mathbf{k}\alpha} = \sqrt{\xi_{\mathbf{k}\alpha}^2 + \Delta^2}$ with corresponding coherence factors which satisfy $u_{\mathbf{k}\alpha}^2 = \frac{1}{2}(1 + \xi_{\mathbf{k}\alpha}/E_{\mathbf{k}\alpha})$, $u_{\mathbf{k}\alpha}^2 + v_{\mathbf{k}\alpha}^2 = 1$. Note that the helicity Green's functions have the standard form, aside from the index $\alpha$.

We define the Nambu spinor $\Psi_{\mathbf{k}}^T = \left( c_{\mathbf{k}\uparrow}, c_{\mathbf{k}\downarrow}, -c_{-\mathbf{k}\downarrow}^\dagger, c_{-\mathbf{k}\uparrow}^\dagger \right)$ of annihilation and creation operators. The many-body Nambu Green's functions are then defined as

$$\mathcal{G}(K) = -\int_0^\beta d\tau e^{i\omega\tau} \left\langle T_\tau \Psi_{\mathbf{k}}(\tau) \Psi_{\mathbf{k}}^\dagger(0) \right\rangle = \begin{pmatrix} G(K) & F(K) \\ \widetilde{F}(K) & \widetilde{G}(K) \end{pmatrix}, \tag{3}$$

where $\beta = 1/T$ is the inverse temperature. The Green's function $G(K)$, time-reversed Green's function $\widetilde{G}(K) = i\sigma_y[G(-K)]^T i\sigma_y$, and the anomalous Green's functions $F(K)$ and $\widetilde{F}(K)$ are now in general $2\times 2$ matrices.

Using the equations of motion for the Nambu Green's function in the pairing approximation we have

$$\mathcal{G}^{-1}(K) = \begin{pmatrix} G_0^{-1}(K) & \Delta \\ \Delta^* & \widetilde{G}_0^{-1}(K) \end{pmatrix}. \tag{4}$$

For the remainder of this supplement, as well as in the main text, we chose $\Delta$ to be real. The non-interacting Green's function of the particle sector is $G_0^{-1}(K) \equiv i\omega - H^0(\mathbf{k})$, and the hole Green's function is defined through $\widetilde{G}_0^{-1}(K) = i\sigma_y[G_0^{-1}(-K)]^T i\sigma_y = i\omega + H^0(\mathbf{k})$. This non-interacting Greens function depends on the single particle Hamiltonian through $H^0(\mathbf{k}) = k^2/2m - \mu + \lambda \mathbf{k}_\perp \cdot \boldsymbol{\sigma}/m$ where $\boldsymbol{\sigma} = (\sigma_x, \sigma_y, \sigma_z)$ is a vector of Pauli matrices in spin space. We consider a 2D Rashba SOC, so that $\boldsymbol{\sigma}$ is coupled only to the in plane momentum $\mathbf{k}_\perp = (k_x, k_y, 0)$. Taking the block matrix inverse gives the following $2 \times 2$ matrix Green's functions:

$$G(K) = \left( G_0^{-1}(K) + \Delta^2 \widetilde{G}_0(K) \right)^{-1}, \tag{5}$$

$$F(K) = \Delta G_0(K) \widetilde{G}(K). \tag{6}$$

In the pairing approximation, the structure of the block matrix inverse forces $\widetilde{F}(K) = \Delta^*/\Delta F(K) = F(K)$.

For pure SOC (i.e., no magnetic field), the system is time-reversal invariant and helicity is a good quantum number. Therefore, the hole Green's function and the particle Green's function can be diagonalized simultaneously with the same unitary transformation: $U_{\mathbf{k}} = \exp\left[ -i\phi_{\mathbf{k}} \pi/4 \left( \hat{\mathbf{k}}_\perp \times \hat{z} \right) \cdot \boldsymbol{\sigma} \right]$ with $\hat{\mathbf{k}}_\perp = \mathbf{k}_\perp/k_\perp$ and $e^{i\phi_{\mathbf{k}}} = (k_x + i k_y)/k_\perp$. This unitary transformation will then diagonalize the non-interacting Green's function $G_0(K)$, and leads to the identity:

$$G_0(K) = \sum_\alpha \frac{1}{i\omega - \xi_{\mathbf{k}\alpha}} P_{\mathbf{k}\alpha}, \tag{7}$$

where $P_{\mathbf{k}\alpha} = U_{\mathbf{k}}(1+\alpha\sigma_z)U_{\mathbf{k}}^\dagger/2 = (1+\alpha \mathbf{k}_\perp \cdot \boldsymbol{\sigma}/k_\perp)/2$ is a projector into the helicity state $\alpha = \pm 1$. We can now express the $G(K)$ and $F(K)$ in terms of the projector $P_{\mathbf{k}\alpha}$ and the helicity Green's functions as Eq. (1) and Eq. (2) through:

$$G(K) = \sum_\alpha G_H^\alpha(K) P_{\mathbf{k}\alpha}, \tag{8}$$

$$F(K) = \sum_\alpha F_H^\alpha(K) P_{\mathbf{k}\alpha}. \tag{9}$$

## II. THE WARD-TAKAHASHI IDENTITY

### A. Deriving the full vertex

The Ward-Takahashi identity (WTI) in quantum field theory is a diagrammatic identity that imposes a symmetry between response functions. The particular symmetry we are interested in is the $U(1)$ abelian gauge symmetry that is present in the density/current channel. Since the pseudogap phase self energy is mean field, the WTI provides a powerful tool in deriving the full vertex that appears in correlation functions. This is because the WTI can be explicitly used to solve for the full vertex function, and thereby obtain the exact correlation function.

We confine our attention here to the normal phase, as the density-density correlation function has collective mode effects which are complicated below $T_c$.

The WTI for the full vertex $\Gamma^\mu(\widetilde{K}, K)$ is[1]

$$q_\mu \Gamma^\mu(\widetilde{K}, K) = G^{-1}(\widetilde{K}) - G^{-1}(K). \tag{10}$$

Similarly the WTI for the bare vertex $\gamma^\mu(\widetilde{K}, K)$ is $q_\mu \gamma^\mu(\widetilde{K}, K) = G_0^{-1}(\widetilde{K}) - G_0^{-1}(K)$ and we also define $q_\mu \widetilde{\gamma}^\mu(\widetilde{K}, K) = \widetilde{G}_0^{-1}(\widetilde{K}) - \widetilde{G}_0^{-1}(K)$. Here $\widetilde{K} \equiv K + Q$. In the pseudogap phase the mean field like self energy is $\Sigma(K) = \Delta^2 \widetilde{G}_0(K)$. Using this self energy, the WTI becomes

$$\begin{aligned} q_\mu \Gamma^\mu(\widetilde{K}, K) &= G_0^{-1}(\widetilde{K}) - G_0^{-1}(K) - \Sigma(\widetilde{K}) + \Sigma(K) \\ &= G_0^{-1}(\widetilde{K}) - G_0^{-1}(K) - \Delta^2 \left( \widetilde{G}_0(\widetilde{K}) - \widetilde{G}_0(K) \right) \\ &= q_\mu \gamma^\mu(\widetilde{K}, K) - \Delta^2 \widetilde{G}_0(\widetilde{K}) \left( \widetilde{G}_0^{-1}(K) - \widetilde{G}_0^{-1}(\widetilde{K}) \right) \widetilde{G}_0(K). \\ &= q_\mu \gamma^\mu(\widetilde{K}, K) - \Delta^2 \widetilde{G}_0(\widetilde{K}) \left( -q_\mu \widetilde{\gamma}^\mu(\widetilde{K}, K) \right) \widetilde{G}_0(K). \end{aligned}$$

It follows that the full vertex is then

$$\Gamma^\mu(\widetilde{K}, K) = \gamma^\mu(\widetilde{K}, K) + \Delta^2 \widetilde{G}_0(\widetilde{K}) \widetilde{\gamma}^\mu(\widetilde{K}, K) \widetilde{G}_0(K). \tag{11}$$

It is important to note that, this expression is exact, and is not a perturbation expansion in terms of bare vertices.

### B. $f$ and longitudinal sum rules

We now show that, given the exact correlation function obtained by using the WTI, the $f$-sum rule is explicitly satisfied. This ensures our correlation functions are consistent with charge conservation.

The correlation function is given by

$$P^{\mu\nu}(Q) = \sum_K \text{Tr} \left[ G(\widetilde{K}) \Gamma^\mu(\widetilde{K}, K) G(K) \gamma^\nu(K, \widetilde{K}) \right]. \tag{12}$$

Inserting the full vertex from Eq. (11) gives the correlation function above $T_c$ as:

$$P^{\mu\nu}(Q) = \sum_K \text{Tr} \left\{ [G(\widetilde{K}) \gamma^\mu(\widetilde{K}, K) G(K) + F_{\text{pg}}(\widetilde{K}) \widetilde{\gamma}^\mu(\widetilde{K}, K) \widetilde{F}_{\text{pg}}(K)] \gamma^\nu(K, \widetilde{K}) \right\}. \tag{13}$$

where $F_{\text{pg}}(K) = \Delta_{\text{pg}} G_0(K) \widetilde{G}(K)$, $\widetilde{F}_{\text{pg}}(K) = \Delta_{\text{pg}} \widetilde{G}_0(K) G(K) = F_{\text{pg}}(K)$.

Applying the WTI to $P^{\mu\nu}(Q)$, and using the cyclic property of the trace, then gives

$$\Omega P^{0\nu}(Q) - \mathbf{q} \cdot P^{i\nu}(Q) = \sum_K \text{Tr} \left\{ G(K) [\gamma^\nu(K, K+Q) - \gamma^\nu(K-Q, K)] \right\}. \tag{14}$$

For the SOC system the bare vertex is $\gamma^\mu(\widetilde{K}, K) = \left(1, \frac{\mathbf{k} + \mathbf{q}/2}{m} + \frac{\lambda}{m} \boldsymbol{\sigma}_\perp \right)$. Using this yields

$$\Omega P^{0\nu}(Q) - \mathbf{q} \cdot P^{i\nu}(Q) = \frac{q^\nu}{m}(1 - \delta_{0,\nu}) \sum_K \text{Tr} \left\{ G(K) \right\} = \frac{nq^\nu}{m}(1 - \delta_{0,\nu}). \tag{15}$$

Here we have used $n = \sum_K \text{Tr}\{G(K)\}$. In components, this equation becomes

$$\omega P^{00}(\omega, \mathbf{q}) - \mathbf{q} \cdot \mathbf{P}^{i0}(\omega, \mathbf{q}) = 0, \tag{16}$$

$$\omega \mathbf{P}^{0j}(\omega, \mathbf{q}) - \mathbf{q} \cdot \overset{\leftrightarrow}{P}{}^{ij}(\omega, \mathbf{q}) = \frac{n}{m}\mathbf{q}. \tag{17}$$

Setting $\omega = 0$ in the second equation and then taking the dot product with $\mathbf{q}$ gives

$$\mathbf{q} \cdot \overset{\leftrightarrow}{P}{}^{ij}(0, \mathbf{q}) \cdot \mathbf{q} = -\frac{nq^2}{m}. \tag{18}$$

Now use the identity $\text{Im} P^{i0}(\omega, \mathbf{q}) = -\text{Im} P^{0i}(-\omega, -\mathbf{q})$ and solve for $\text{Im} P^{00}$ in terms of $\text{Im}\overset{\leftrightarrow}{P}{}^{ij}$. Applying the Kramers-Kronig relations and Eq. (18) then gives

$$\int \frac{d\omega}{\pi}(-\omega \text{Im} P^{00}(\omega, \mathbf{q})) = \int \frac{d\omega}{\pi}\left(-\frac{\mathbf{q} \cdot \text{Im}\overset{\leftrightarrow}{P}{}^{ij}(\omega, \mathbf{q}) \cdot \mathbf{q}}{\omega}\right)$$
$$= -\mathbf{q} \cdot \text{Re}\overset{\leftrightarrow}{P}{}^{ij}(\mathbf{q}, 0) \cdot \mathbf{q}$$
$$= \frac{nq^2}{m}. \tag{19}$$

Defining $\chi_{\rho\rho}(Q) \equiv P^{00}(Q), \overset{\leftrightarrow}{\chi}_{JJ}(Q) \equiv P^{ij}(Q), i, j \in \{1, 2, 3\}$, the $f$-sum rule, which holds for all $\mathbf{q}$, is thus

$$\int \frac{d\omega}{\pi}\left(-\omega \chi''_{\rho\rho}(\omega, \mathbf{q})\right) = \frac{nq^2}{m}. \tag{20}$$

Here the singular part of the correlation function, $\chi''_{\rho\rho}(\omega, \mathbf{q})$, is equal to $\text{Im}\chi_{\rho\rho}(\omega, \mathbf{q})$, which is true only in the density/current channel. The next section on the spin-spin correlation functions elaborates further on this point. Similarly the longitudinal sum rule, which holds for all $\mathbf{q}$ in this continuum case, and above $T_c$, is thus

$$\int \frac{d\omega}{\pi}\left(-\frac{\mathbf{q} \cdot \overset{\leftrightarrow}{\chi}''_{JJ}(\omega, \mathbf{q}) \cdot \mathbf{q}}{\omega}\right) = \frac{nq^2}{m}. \tag{21}$$

## III. SPIN-SPIN CORRELATION FUNCTIONS AND SUM RULES

### A. General sum rules for spin density-spin density correlation functions

Here we investigate both the spin-spin correlation functions and sum rules. In the presence of SOC, spin is no longer conserved. The WTI is therefore inapplicable to deduce the form of the spin-spin correlation functions. However, the Heisenberg equations of motion can be utilized to derive a sum rule for spin density-spin density correlation functions.

We start with the definition for the spin-spin correlation function in the main text, $\chi_{S_i S_j}(i\omega, \mathbf{q}) \equiv \int d\tau\, e^{i\omega\tau} \langle T_\tau S_{\mathbf{q}i}(\tau) S_{-\mathbf{q}j}(0)\rangle$, for a many-body spin operator $S_{\mathbf{q}i} = \sum_{ss'\mathbf{k}} c^\dagger_{\mathbf{k}s}(\sigma_i)_{ss'} c_{\mathbf{k}+\mathbf{q}s'}$. We will consider a general many-body Hamiltonian $\mathcal{H} = \mathcal{H}_0 + \mathcal{H}_I$, where the contribution $\mathcal{H}_0 = \sum_{ss'\mathbf{k}} c^\dagger_{\mathbf{k}s} H^0_{ss'}(\mathbf{k}) c_{\mathbf{k}s'}$ is constructed from an arbitrary single particle Hamiltonian $H^0_{ss'}(\mathbf{k})$. We also assume that the interaction Hamiltonian commutes with the spin-operator, $[\mathcal{H}_I, S_{\mathbf{q}i}] = 0$.

In order to constrain the spin-spin correlation function, we investigate the analogous sum rule for $\chi''_{S_i S_j}(\omega, \mathbf{q})$ that is derived in the previous section for the density/current response. To do this, we first analytically continue $\chi_{S_i S_j}(i\omega, \mathbf{q})$ to real frequency by setting $i\omega = \omega + i\delta$, then letting $\delta \to 0$, and finally invoking the Sokhotski–Plemelj theorem[2]. This is expressed mathematically as

$$\chi_{S_i S_j}(\omega, \mathbf{q}) \equiv \lim_{\delta \to 0} \chi_{S_i S_j}(i\omega, \mathbf{q})\Big|_{i\omega = \omega + i\delta},$$
$$= \chi'_{S_i S_j}(\omega, \mathbf{q}) + i\chi''_{S_i S_j}(\omega, \mathbf{q}),$$
$$= \mathcal{P} \int \frac{dz}{\pi} \frac{\chi''_{S_i S_j}(z, \mathbf{q})}{z - \omega} + i\chi''_{S_i S_j}(\omega, \mathbf{q}). \tag{22}$$

Here $\mathcal{P}$ denotes the Cauchy principal value. Note that, for $i \neq j$, $\chi''_{S_i S_j}(\omega, \mathbf{q})$ need not necessarily be real[3]. Returning to the time domain, the singular response function is $\chi''_{S_i S_j}(t, \mathbf{q}) = \frac{1}{2}\langle [S_{\mathbf{q}i}(t), S_{-\mathbf{q}j}(0)]\rangle$.

We now can derive the sum rule. Using the properties of Fourier transforms, we have

$$\int \frac{d\omega}{\pi} \left(-\omega \chi''_{S_i S_j}(\omega, \mathbf{q})\right) = -2i\partial_t \chi''_{S_i S_j}(t, \mathbf{q})\Big|_{t=0},$$
$$= -\langle [i\partial_t S_{\mathbf{q}i}(t), S_{-\mathbf{q}j}(0)]\rangle|_{t=0},$$
$$= \langle [[\mathcal{H}, S_{\mathbf{q}i}], S_{-\mathbf{q}j}]\rangle, \quad (23)$$

where we have used the Heisenberg equations of motion $i\partial_t S_{\mathbf{q}i}(t) = -[\mathcal{H}, S_{\mathbf{q}i}(t)]$ and the commutators on the last line are taken at equal time. The problem thus reduces to calculating the equal time commutator $\langle [[\mathcal{H}, S_{\mathbf{q}i}], S_{-\mathbf{q}j}]\rangle = [\mathcal{H}_0, S_{\mathbf{q}i}]$.

The commutator in Eq. (23) is tedious but straightforward to calculate. After much algebra we arrive at the $f$-sum rule, or analogue thereof, for spin density-spin density response:

$$\int \frac{d\omega}{\pi} \left(-\omega \chi''_{S_i S_j}(\omega, \mathbf{q})\right) = \frac{1}{2} \sum_{\mathbf{k}ss'} \bigg( \{\sigma_i, H^0(\mathbf{k}) - H^0(\mathbf{k}+\mathbf{q})\}\sigma_j - \sigma_j\{\sigma_i, H^0(\mathbf{k}-\mathbf{q}) - H^0(\mathbf{k})\}$$
$$+ \left[H^0(\mathbf{k}) + H^0(\mathbf{k}+\mathbf{q}), \sigma_i\right]\sigma_j - \sigma_j\left[H^0(\mathbf{k}-\mathbf{q}) + H^0(\mathbf{k}), \sigma_i\right]\bigg)_{ss'} \langle c^\dagger_{\mathbf{k}s} c_{\mathbf{k}s'}\rangle. \quad (24)$$

This expression can be further simplified as follows. The Hamiltonian can be expressed as $H^0(\mathbf{k}) = h_\mu(\mathbf{k})\sigma^\mu$ where $h_\mu(\mathbf{k}) = \frac{1}{2}\text{Tr}\left[\sigma_\mu H^0(\mathbf{k})\right]$ and $\sigma^\mu = (1, \boldsymbol{\sigma})$. After using the commutation and anti-commutation rules for Pauli matrices, further algebra reduces the sum rule to the form

$$\int \frac{d\omega}{\pi}\left(-\omega \chi''_{S_i S_j}(\omega, \mathbf{q})\right) = \sum_{\mathbf{k}} \bigg\{ \text{Tr}\left[\left(H^0(\mathbf{k}+\mathbf{q}) + H^0(\mathbf{k}-\mathbf{q}) - 2H^0(\mathbf{k})\right)g_{\mathbf{k}}\right]\delta_{ij}$$
$$- 2\left((\mathbf{h}(\mathbf{k}+\mathbf{q}) + \mathbf{h}(\mathbf{k}-\mathbf{q}))\cdot \mathbf{s}_{\mathbf{k}}\delta_{ij} + (h_i(\mathbf{k}+\mathbf{q}) + h_i(\mathbf{k}-\mathbf{q}))s_{\mathbf{k}j}\right)$$
$$- i\left[(h_k(\mathbf{k}+\mathbf{q}) - h_k(\mathbf{k}-\mathbf{q}))n_{\mathbf{k}} - (h_0(\mathbf{k}+\mathbf{q}) - h_0(\mathbf{k}-\mathbf{q}))s_{\mathbf{k}k}\right]\epsilon_{ijk}\bigg\}, \quad (25)$$

where we have defined $g_{\mathbf{k}} = T\sum_{i\omega} G(i\omega, \mathbf{k})$, $n_{\mathbf{k}} = \text{Tr}[g_{\mathbf{k}}]$ and $s_{\mathbf{k}i} = \text{Tr}[\sigma_i g_{\mathbf{k}}]$. The first line, appearing only for $i = j$, can be seen to be equal to the density $f$-sum rule contribution. The second line can give diagonal contributions, resulting in a deviation of the spin density response from the density response in the presence of SOC. The third line vanishes for $i = j$. It is of note that this line is purely imaginary, and can give an imaginary $f$-sum contribution. This is only possible for correlation functions of two distinct operators, as was noted in the literature[3]. This implies that $\chi''_{S_i S_j} \neq \text{Im}\chi_{S_i S_j}$ for $i \neq j$. Finally, the first and third lines of Eq. (25) are symmetric under interchange of $i \leftrightarrow j$ and $\mathbf{q} \leftrightarrow -\mathbf{q}$, while the second line is not. This can be explained by observing that Eq. (23) is not fully symmetric under the the exchange of $i \leftrightarrow j$ and $\mathbf{q} \leftrightarrow -\mathbf{q}$.

### B. Explicit spin-spin correlation functions and sum rules for Rashba SOC

We now present simplified expressions for the correlation functions and sum rules for the case of Rashba SOC. Due to symmetry in the $x$-$y$ plane, there are only four unique correlation functions of interest. The remaining spin density correlation functions can be found through exchanging $x \to y$ where relevant. The explicit spin-spin correlation functions are

$$\chi_{S_x S_x}(\omega, \mathbf{q}) = \frac{1}{2}\sum_{K,\alpha,\alpha'}(1 + \alpha\alpha'\cos(\phi_{\mathbf{k}+\mathbf{q}} + \phi_{\mathbf{k}}))\left[G^\alpha_H(K)G^{\alpha'}_H(\widetilde{K}) + F^\alpha_H(K)F^{\alpha'}_H(\widetilde{K})\right], \quad (26)$$

$$\chi_{S_x S_z}(\omega, \mathbf{q}) = \frac{i}{2}\sum_{K,\alpha,\alpha'}(\alpha\alpha'\sin(\phi_{\mathbf{k}+\mathbf{q}}) - \sin(\phi_{\mathbf{k}}))\left[G^\alpha_H(K)G^{\alpha'}_H(\widetilde{K}) + F^\alpha_H(K)F^{\alpha'}_H(\widetilde{K})\right], \quad (27)$$

$$\chi_{S_z S_z}(\omega, \mathbf{q}) = \frac{1}{2}\sum_{K,\alpha,\alpha'}(1 - \alpha\alpha'\cos(\phi_{\mathbf{k}+\mathbf{q}} - \phi_{\mathbf{k}}))\left[G^\alpha_H(K)G^{\alpha'}_H(\widetilde{K}) + F^\alpha_H(K)F^{\alpha'}_H(\widetilde{K})\right], \quad (28)$$

$$\chi_{S_x S_y}(\omega, \mathbf{q}) = 0. \quad (29)$$

Note that the response $\chi_{S_x S_y}$ vanishes identically, as it must due to in-plane rotational symmetry.

For each of the above correlation functions, we have explicitly integrated the left hand side of the sum rule, Eq. (25), and found:

$$\int \frac{d\omega}{\pi} \left(-\omega \chi''_{S_x S_x}(\omega, \mathbf{q})\right) = \frac{nq^2}{m} - \frac{4\lambda}{m} \sum_{\mathbf{k}\alpha} \alpha \frac{k_x^2}{k_\perp} n_{\mathbf{k}\alpha}, \quad (30)$$

$$\int \frac{d\omega}{\pi} \left(-\omega \chi''_{S_x S_z}(\omega, \mathbf{q})\right) = -iq_y \frac{4\lambda}{m} \sum_{\mathbf{k}\alpha} \left(\alpha \frac{k_y^2}{k_\perp} - \lambda\right) n_{\mathbf{k}\alpha}, \quad (31)$$

$$\int \frac{d\omega}{\pi} \left(-\omega \chi''_{S_z S_z}(\omega, \mathbf{q})\right) = \frac{nq^2}{m} - \frac{4\lambda}{m} \sum_{\mathbf{k}\alpha} \alpha k_\perp n_{\mathbf{k}\alpha}, \quad (32)$$

where $n_{\mathbf{k}\alpha} = \text{Tr}[P_{\mathbf{k}\alpha} g_\mathbf{k}] = T \sum_{i\omega} G_H^\alpha(K)$. These results are consistent with the form using the right hand side of Eq. (25). Note further that the sum rule for $\chi_{S_x S_z}$ is purely imaginary. This is not surprising, as mentioned above.

## IV. SUPERFLUID DENSITY

The formalism underlying this paper is capable of addressing the superfluid phase. However, many response functions in this phase require the inclusion of collective mode effects. The superfluid density, which can be obtained using a transverse response, does not require collective modes. The superfluid density $\left(\frac{\overleftrightarrow{n}_s}{m}\right)$ is defined by

$$\left(\frac{\overleftrightarrow{n}_s}{m}\right) \equiv \left(\frac{\overleftrightarrow{n}}{m}\right)_{\text{dia}} - \overleftrightarrow{\chi}_{JJ}^T(0), \quad (33)$$

where $\overleftrightarrow{\chi}_{JJ}^T(Q)$ is the transverse part of the current-current correlation function.

This contains two contributions from: ($i$) quasi-particles which are reflected in the current-current correlation function and ($ii$) the diamagnetic current which above $T_c$ must precisely cancel the quasi-particle term. Below $T_c$ the current-current correlation function acquires an addition contribution which depends on the anomalous Green's function $F_{\text{sc}}$. Note that the diamagnetic current is dependent on the total pairing gap. The current-current correlation function depends on the function $F_{\text{pg}}$ as defined above, and arises from vertex corrections.

Using Eq. (10) from the main paper, the current-current correlation function and diamagnetic response tensors are

$$\overleftrightarrow{\chi}_{JJ}(Q) = \sum_K \text{Tr}\left[\mathbf{J}(\mathbf{k},\mathbf{q}) G(\widetilde{K}) \mathbf{J}(\mathbf{k},\mathbf{q}) G(K) + \mathbf{J}(\mathbf{k},\mathbf{q}) F_{\text{pg}}(\widetilde{K}) \mathbf{J}(\mathbf{k},\mathbf{q}) \widetilde{F}_{\text{pg}}(K) - \mathbf{J}(\mathbf{k},\mathbf{q}) F_{\text{sc}}(\widetilde{K}) \mathbf{J}(\mathbf{k},\mathbf{q}) \widetilde{F}_{\text{sc}}(K)\right], \quad (34)$$

$$\left(\frac{\overleftrightarrow{n}}{m}\right)_{\text{dia}} = \sum_K \text{Tr}\left[\mathbf{J}(\mathbf{k},0) G(K) \mathbf{J}(\mathbf{k},0) G(K) + \mathbf{J}(\mathbf{k},0) F_{\text{pg}}(K) \mathbf{J}(\mathbf{k},0) \widetilde{F}_{\text{pg}}(K) + \mathbf{J}(\mathbf{k},0) F_{\text{sc}}(K) \mathbf{J}(\mathbf{k},0) \widetilde{F}_{\text{sc}}(K)\right], \quad (35)$$

where $\mathbf{J}(\mathbf{k},\mathbf{q}) = \left(\frac{\mathbf{k}+\mathbf{q}/2}{m} + \frac{\lambda}{m}\boldsymbol{\sigma}_\perp\right)$ is the current operator. Notice that the relative sign between the $F_{\text{sc}}$ and $F_{\text{pg}}$ contributions is different in $\overleftrightarrow{\chi}_{JJ}^T(Q)$ and in the diamagnetic current.

The superfluid density tensor is thus

$$\left(\frac{\overleftrightarrow{n}_s}{m}\right)_{ij} = 2\sum_K \text{Tr}\left[J_i(\mathbf{k},0) F_{\text{sc}}(K) J_j(\mathbf{k},0) \widetilde{F}_{\text{sc}}(K)\right]. \quad (36)$$

Due to symmetry, the off-diagonal ($i \neq j$) elements vanish. Furthermore, the SOC leads to different in-plane ($i = j = x, y$) and out-of-plane ($i = j = z$) response. Despite a significantly different form, we can show this expression for the superfluid density is consistent with that derived by adding a phase twist to the order parameter[4-6].

---



\end{document}